\documentclass[prb,twocolumn,superscriptaddress,showpacs,citeautoscript]{revtex4}

\usepackage{graphicx}
\usepackage{amsmath}
\usepackage{amssymb}

\pdfoutput=1

\begin{document}
	
\title[Spin-dependent thermoelectric properties of a Kondo-correlated quantum dot ]{Spin-dependent thermoelectric properties of a Kondo-correlated quantum dot with Rashba spin-orbit coupling}

\author{\L{}ukasz Karwacki}
\email{karwacki@amu.edu.pl}
\affiliation{Faculty of Physics, Adam Mickiewicz University, 61-614 Pozna\'{n}
Poland}
\author{Piotr Trocha}
\affiliation{Faculty of Physics, Adam Mickiewicz University, 61-614 Pozna\'{n}
Poland}
\author{J\'{o}zef Barna\'{s}}
\affiliation{Faculty of Physics, Adam Mickiewicz University, 61-614 Pozna\'{n}
Poland}
\affiliation{Institute of Molecular Physics, Polish Academy of Sciences,
60-179 Pozna\'{n}, Poland}

\begin{abstract}
Thermoelectric transport phenomena in a single-level quantum dot coupled to ferromagnetic leads are considered theoretically in the Kondo regime.
The dot is described by the Anderson model with Rashba type spin-orbit interaction. The finite-U mean field slave boson technique is used to describe the  transport characteristics, like heat conductance, thermopower, thermoelectric efficiency (figure of merit). The role of quantum interference effects in thermoelectric parameters is also analyzed.
\end{abstract}

\pacs{ 73.23.-b, 73.63.Kv\, 72.15.Qm, 85.35.Ds}

\maketitle
\section{Introduction}\label{Sec:1}

Various quantum interference effects, like Dicke and Fano phenomena~\cite{dicke1,dicke2,Fano} are well known in atomic physics and quantum optics. These phenomena have been reported also in electronic transport through mesoscopic systems~\cite{guevaraD,shahbazyan,brandes,trocha,trocha1}.
Originally, the Dicke effect was observed in spontaneous atomic emission spectra as a strong and very narrow resonance which coexisted with a much broader line. The phenomenon occurs only if the distance between atoms is much smaller than the wavelength of
the emitted light (by an individual atom). The broad resonance, associated with a state that is strongly coupled to the electromagnetic field, is called superradiant mode, whereas the narrow peak can be attributed to a state which is weakly  coupled to the electromagnetic field and is referred to
as the subradiant mode.

In electronic transport through quantum dots, in turn, this effect is due
to indirect coupling of QDs through the leads~\cite{shahbazyan}.
Although, the electronic version of Dicke effect has been studied theoretically in many papers~\cite{guevaraD,shahbazyan,brandes,trocha,trocha1,ulloa,trochaJNN} there is still no experimental verification of the phenomenon.

The Fano effect, on the other hand, reveals in experiment as an asymmetric line in emission
spectra. It is caused by quantum interference of waves resonantly
transmitted through a discrete level and those transmitted
nonresonantly through a continuum of states~\cite{Fano}.
For example, the Fano antiresonance  may be identified in photoelectron ionization spectra of an autoionization system with one discrete level interacting with a neighbor two-level atom~\cite{leonski}.
The Fano effect is well known not only in optics, but also in mesoscopic physics.
It has been observed in systems consisting of a quantum dot attached to external leads. The discrete dot level serves then as a resonant channel~\cite{Gores2,clerk,kobayashi,johnson,sasaki}, and the interference of electrons transmitted through the resonant channel and those tunneling nonresonantly directly between the leads gives rise to the Fano effect. The effect can be also observed in a double quantum dot system due to asymmetry in the coupling strengths of the bonding and antibonding states to the external electrodes~\cite{sasaki,guevara,lu05,trocha2,kuboPRB06}.

To observe the above described resonance phenomena one needs two different and coherent transport channels. In the case of a double quantum dot system, this can be achieved when there is a coherent indirect tunneling of electrons between the dots through the leads. As far as we need only two different channels, one can use a single-level quantum dot with spin-polarized transport. This can be realized when the dot is coupled to external ferromagnetic leads~\cite{hamaya1,hamaya2,hamaya3}, and the two channels correspond to two electron spin orientations. However, without introducing additional interactions, these channels remain independent and no interference effects can occur. The interference appears when there is some mixing of the two spin channels, caused  for instance  by Rashba spin-orbit interaction. Such a Rashba-induced Fano resonance has been theoretically predicted in spin-polarized transport through a quasi-one-dimensional spin transistor~\cite{datta90}. The effect appears then  due to the spin-orbit induced mixing of the continuum
and bound states in the region between the ferromagnetic leads~\cite{shelykh04}. More recently, Rashba-induced Fano resonance has been investigated in electronic transport through a quantum wire with localized Rashba interaction~\cite{sanchez06,lopezR07,Crisan09}.

Quantum interference effects can significantly modify electrical conductance of the system.
But the impact of these phenomena on transport properties is much broader. The interference effects can also modify thermoelectric coefficients~\cite{liuJAP,trocha12Th,karlstrom}. More specifically, the resonant transport may significantly modify the thermal conductance as well as the thermopower and thermoelectric efficiency (figure of merit $ZT$).
Recently, the thermoelectric properties of nanostructures have attracted great interest due to enhanced thermoelectric efficiency associated with quantum confinement and Coulomb blockade effects~\cite{hick,mahan,beenak,blanter,turek,koch,kubala1,zianni,zhangxm}.
Moreover, the lattice thermal conductance of low dimensional systems is
rather small~\cite{hochbaum,baheti,boukai}, which also can  lead to a relatively high thermoelectric efficiency.
This applies not only to the conventional thermoelectric effects, but also to their spin counterparts.

In this paper the considerations are focused on the impact of interference effects on thermoelectric phenomena.  We consider a single-level quantum dot coupled to ferromagnetic leads. This coupling includes both spin-conserving and spin-flip tunneling processes -- the latter due to Rashba spin-orbit interaction.  It is already well known, that ferromagnetism of the leads has a significant influence on the thermoelectric properties of a single-level quantum dot~\cite{dubiPRB09,swirkowicz}.
Here, the considerations are limited to the  Kondo regime~\cite{kondo,cronewett,gores,glazman}, where spin correlations can strongly influence
the  thermoelectric transport properties~\cite{boese,dong,krawiec,sakano,kim,scheibner,franco,yoshida,costi,zitkoPRB12}.
To describe electronic transport through the considered system at low temperatures, we apply the finite-$U$ slave boson mean field approach introduced in~\cite{kotliar}.
Accordingly, we study thermoelectric properties of the system at temperatures  below the  Kondo temperature.
More specifically, we calculate the electronic part of the thermal conductance, thermoelectric efficiency (figure of merit), Seebeck coefficient, and the spin Seebeck coefficient.

The paper is organized as follows. In Section 2 we describe the
 quantum dot system  under consideration.
We also briefly describe there the finite-$U$ slave-boson mean-field technique used to
calculate the basic transport characteristics. Numerical results
 are presented and discussed in Section 3.
Final conclusions are included in Section 4.

\section{Theoretical description}\label{Sec:2}

\subsection{Model}

We consider a single-level quantum dot coupled to external ferromagnetic leads
by spin-conserving and spin-nonconserving interactions. Apart from this, magnetic moments of both electrodes are assumed to be parallel.
The whole system can be described by  Hamiltonian of the form
\begin{eqnarray}\label{hamiltonian}
H=H_{e}+H_{QD}+H_{t}+H_{t}^{so}.
\end{eqnarray}
The term $H_{e}=\sum_{\textbf{k}\beta\sigma}\varepsilon_{\textbf{k}\beta\sigma}c_{\textbf{k}\beta\sigma}^{\dagger}c_{\textbf{k}\beta\sigma}$
describes  Bloch electrons in the left and right ferromagnetic electrodes ($\beta =L,R)$, corresponding to the wave vector \textbf{k} and spin $\sigma(=\uparrow,\downarrow)$.
The next term in~\ref{hamiltonian} describes the isolated quantum dot and takes the form
\begin{eqnarray}
H_{QD}=\sum_{\sigma}\varepsilon_{\sigma}d_{\sigma}^{\dagger}d_{\sigma}+Ud_{\uparrow}^{\dagger}d_{\uparrow}d_{\downarrow}^{\dagger}d_{\downarrow}.
\end{eqnarray}
Here $\varepsilon_{\sigma}=\varepsilon_{d}-\hat{\sigma}B$, where $\varepsilon_{d}$ is the energy of spin degenerate dot's level, which can be controlled by an external gate voltage, $B$ is a magnetic field measured in energy units ($B$ is along magnetic moments of the leads), while $\hat{\sigma}=1(-1)$
for spin up (spin down) electrons. The parameter $U$ is the Coulomb energy of two electrons occupying the dot's level.

The last two terms in~\ref{hamiltonian}, $H_{t}$ and $H_{t}^{so}$,  stand for electron tunneling processes between the dot and the external leads.  The component $H_{t}$ represents the tunneling processes that conserve electron spin, and acquires the standard form,
\begin{eqnarray}
H_{t}=\sum_{\textbf{k}\beta\sigma}V_{\textbf{k}\beta\sigma}c_{\textbf{k}\beta\sigma}^{\dagger}d_{\sigma}+{\rm h.c.}
\end{eqnarray}
 In turn, the component $H_{t}^{so}$ describes tunneling processes which do not conserve the electron spin.
This term, assumed here in the form of  Rashba-type  spin-orbit coupling, leads to mixing of the electron states
of opposite  spin orientations (as defined in the electrodes). We write this term in the
form~\cite{orellanaNano,key-1}
\begin{eqnarray}\label{rashba_ham}
H_{t}^{so}=-\sum_{\textbf{k}\beta\sigma}\left(V_{\textbf{k}\beta\overline{\sigma}}^{so}c_{\textbf{k}\beta\overline{\sigma}}^{\dagger}
(i\mathbf{\sigma}_{x})_{\sigma\overline{\sigma}}d_{\sigma}+h.c.\right),
\end{eqnarray}
where $V_{\textbf{k}\beta\sigma}^{so}$ is the parameter of Rashba spin-orbit coupling, while $\overline{\sigma}=\downarrow$ for $\sigma=\uparrow$ and
$\overline{\sigma}=\uparrow$ for $\sigma=\downarrow$.
In the following we assume $V_{\bf{k}\beta\sigma}$ to be independent of $\bf{k}$ and introduce the coupling parameter $\Gamma_{\beta\sigma}$ defined as
$\Gamma_{\beta\sigma}=2\pi |V_{\beta\sigma}|^2 \rho_{\beta\sigma}$, where $\rho_{\beta\sigma}$ is the spin dependent density of states in the lead $\beta$.
Similarly, we also define $\Gamma_{\beta\sigma}^{so}=2\pi |V_{\beta\sigma}^{so}|^2 \rho_{\beta\sigma}$.
Furthermore, we write
$\Gamma_{\beta\sigma}=(1\pm\hat{\sigma} p_{\beta})\Gamma_{\beta}$, where $p_\beta$ is the spin polarization of the lead $\beta$ and $\Gamma_{\beta}$ is the coupling parameter. Similarly we introduce $\Gamma_{\beta\sigma}^{so}=(1\pm\hat{\sigma} p_{\beta})\Gamma^{so}_{\beta}$,  and write
$\Gamma_{\beta}^{so}=q\Gamma_{\beta}$, where q is a parameter which describes relative strength of the spin-orbit coupling.

From the experimental point of view, such a system could be realized by making use of quantum
point contacts to connect the quantum dot to electrodes. We note that the interplay of
spin-orbit interaction and magnetic field leads to the interference
Fano effect which has already been described in the Coulomb regime~\cite{stefanskiJP}.
Furthermore, the Fano effect in quantum dot structures is known to suppress both the
electronic and heat conductances, and to enhance the thermoelectric current
generation~\cite{trocha12Th}.

The above model is equivalent to a spinless two-level quantum dot with an effective magnetic field noncollinear with magnetization of the leads and coupled to a pseudo-spin. Such a model was studied recently in the context of phase-lapses and population inversion in the Kondo regime~\cite{kashcheyevs, silvestrov}. The equivalence of the two models is shown in the Appendix, where a relevant canonical transformation is used as described in Ref. [ \onlinecite{kashcheyevs}].

\subsection{Method}

To describe the Kondo correlations in the system under consideration we use
the well known bosonization scheme introduced by Kotliar and R\"uckenstein~\cite{kotliar}.
This method has been shown to describe accurately correlations in
quantum dot systems~\cite{key-4,key-5} -- both in the linear
and nonlinear (but not too far from equilibrium) transport regimes.

In the slave boson mean field theory (SBMFT) \cite{kotliar}, standard fermion operators for electrons in the quantum dot
are replaced with pseudo-fermion operators $d_{\sigma}^{(\dagger)}\rightarrow f_{\sigma}^{(\dagger)}z_{\sigma}^{(\dagger)}$,
where $z_{\sigma}=e^{\dagger}p_{\sigma}+p_{\overline{\sigma}}^{\dagger}d$
is a projection operator describing the many-body effects that
accompany the annihilation of an electron in the dot. Henceforth, it is assumed that the operator
$z_{\sigma}^{(\dagger)}$ and the bosonic field operators $e^{(\dagger)},p_{\sigma}^{(\dagger)},d^{(\dagger)}$
can be approximated by their mean values $z_{\sigma}$,  $e$, $p_{\sigma}$, and $d$.
To conserve Fock's space of the quantum dot, some constraints
have to be imposed. The first constraint, $\sum_{\sigma}p_{\sigma}^{2}+d^{2}+e^{2}-1=0$,
is the state conservation. The second constraint, $p_{\sigma}^{2}+d^{2}-(1/2\pi)\int\, d\varepsilon\, G_{\sigma}^{<}=0$,
is the charge conservation. Here, $G_{\sigma}^{<}$ is the lesser (Keldysh) Green function.
 All this leads to the effective Hamiltonian which differs from the original
one. First, the quantum dot Hamiltonian takes the form
$\tilde{H}_{QD}=\sum_{\sigma}\tilde{\varepsilon}_{\sigma}f_{\sigma}^{\dagger}f_{\sigma}$,
where $\tilde{\varepsilon}_{\sigma}=\varepsilon_{\sigma}+\lambda_{\sigma}^{(2)}$.
The Coulomb interaction term in the SBMFT has no longer the operator form, and appears along with the boundary conditions as the additional term   $E_{g}=Ud^{2}+\lambda^{(1)} \left( e^{2}+\sum_{\sigma}p_{\sigma}^{2}+d^{2}-1 \right) -
\sum_{\sigma}\lambda_{\sigma}^{(2)} \left( p_{\sigma}^{2}+d^{2} \right)$ in the Hamiltonian.
Second, $\Gamma_{\beta\sigma}$ and $\Gamma_{\beta\sigma}^{so}$ are
modified to include the effective bosonic field term as follows; $\tilde{\Gamma}_{\beta\sigma}= \Gamma_{\beta\sigma}z_{\sigma}^{2}$ and $\tilde{\Gamma}_{\beta\sigma}^{so}= \Gamma_{\beta\sigma}^{so}z_{\sigma}^{2}$.

Using Hellman-Feynman theorem,
$\partial\langle H_{eff}\rangle/\partial \chi=0$
(where $\chi$ is one of the parameters: $\lambda^{(1)}, \lambda_{\sigma}^{(2)}, e, p_{\sigma}, d$),
one arrives at the following system of nonlinear equations~\cite{key-4,key-5}:
\begin{eqnarray}\label{sc_a}
\sum_{\sigma}\frac{\partial\ln(z_{\sigma})}{\partial e}\int\frac{d\varepsilon}{2\pi}(\varepsilon-\tilde{\varepsilon}_{\sigma})G_{\sigma}^{<}+\lambda^{(1)}e=0,
\end{eqnarray}
\begin{eqnarray}
\sum_{\sigma}\frac{\partial\ln(z_{\sigma})}{\partial p_{\sigma'}}\int\frac{d\varepsilon}{2\pi}(\varepsilon-\tilde{\varepsilon}_{\sigma})G_{\sigma}^{<}+\left(\lambda^{(1)}-\lambda_{\sigma'}^{(2)}\right)p_{\sigma'}\\ \nonumber \ =0, : \hskip 0.5cm {\rm for} \: \sigma'=\uparrow,\downarrow
\end{eqnarray}
\begin{eqnarray}\label{sc_c}
\sum_{\sigma}\frac{\partial\ln(z_{\sigma})}{\partial d}\int\frac{d\varepsilon}{2\pi}(\varepsilon-\tilde{\varepsilon}_{\sigma})G_{\sigma}^{<}+\left(U+\lambda^{(1)}-\sum_{\sigma}\lambda_{\sigma}^{(2)}\right)d \\ \nonumber  =0.
\end{eqnarray}
These equations, together with the constraints described above, have to be solved self-consistently. To find numerical solutions to these equations, we need to know the corresponding Green functions, which are matrices in the spin space.

Since  the Rashba term~(\ref{rashba_ham}) in the Hamiltonian leads to mixing of the spin components,
the off-diagonal elements of the Green's functions
matrix in the spin space are nonzero.
The retarded Green's
function $\textbf{G}^{r}$ can be derived making use of the  Dyson's
equation, which allows to write $\textbf{G}^{r}$  as $\textbf{G}^{r}=\begin{pmatrix}\textbf{g}_{0}^{r}-{\mathbf{\Sigma}}^{r}\end{pmatrix}^{-1}$,
where $g_{0\sigma\sigma'}^{r}=\delta_{\sigma\sigma'}\left(\varepsilon-\tilde{\varepsilon}_{\sigma}+i0^{+}\right)^{-1}$
is the Green function of the dot decoupled from the leads,
and ${\mathbf{\Sigma}}^{r}={\mathbf{\Sigma}}_{L}^{r}+{\mathbf{\Sigma}}_{R}^{r}$ is the self-energy matrix where the $\beta$ component has
the form
\begin{widetext}
\begin{eqnarray}\label{coupl}
\mathbf{\Sigma}_{\beta}^{r}=
\begin{pmatrix}
\frac{i}{2}\left(\tilde{\Gamma}_{\beta\uparrow}+\tilde{\Gamma}_{\beta\downarrow}^{so}\right) &
\frac{1}{2}\left(\sqrt{\tilde{\Gamma}_{\beta\uparrow}\tilde{\Gamma}_{\beta\uparrow}^{so}}
-\sqrt{\widetilde{\Gamma}_{\beta\downarrow}\widetilde{\Gamma}_{\beta\downarrow}^{so}}\right) \\
-\frac{1}{2}\left(\sqrt{\tilde{\Gamma}_{\beta\uparrow}\tilde{\Gamma}_{\beta\uparrow}^{so}}
-\sqrt{\widetilde{\Gamma}_{\beta\downarrow}\widetilde{\Gamma}_{\beta\downarrow}^{so}}\right) &
\frac{i}{2}\left(\tilde{\Gamma}_{\beta\downarrow}+\tilde{\Gamma}_{\beta\uparrow}^{so}\right)
\end{pmatrix},
\end{eqnarray}
\end{widetext}
for $\beta =L,R$.
Having found the retarded and advanced Green functions, one can find the lesser (Keldysh)  Green function
from the formula    $\mathbf{G}^{<}=i(f_{L}\mathbf{G}^{r}\tilde{\mathbf{\Gamma}}_{L}\mathbf{G}^{a}+f_{R}\mathbf{G}^{r}\tilde{\mathbf{\Gamma}}_{R}\mathbf{G}^{a})$.
Here, $\tilde{\mathbf{\Gamma}}_{L(R)}$ is defined as
\begin{eqnarray}
\tilde{\mathbf{\Gamma}}_{L(R)}=-i\mathbf{\Sigma}_{L(R)}.
\end{eqnarray}
The Green function  $G^<_\sigma$ in Eqs~(\ref{sc_a})-(\ref{sc_c}) is the diagonal element of $\mathbf{G}^{<}$.
Finally, one can calculate the transmission function $T(\varepsilon)$ as
\begin{eqnarray}
T(\varepsilon)= {\rm Tr}\left( \mathbf{G}^{a}\tilde{\mathbf{\Gamma}}_{R}\mathbf{G}^{r}\tilde{\mathbf{\Gamma}}_{L} \right).
\end{eqnarray}
Note, the maximum value of the transmission function $T(\varepsilon)$ is equal to the number of quantum transport channels. In addition, the
transmission function $T(\varepsilon)$ can be written as $T(\varepsilon)= \sum_\sigma T_\sigma (\varepsilon)$, where
$T_\sigma (\varepsilon)$ are the diagonal elements of $ \mathbf{G}^{a}\tilde{\mathbf{\Gamma}}_{R}\mathbf{G}^{r}\tilde{\mathbf{\Gamma}}_{L}$.

\subsection{Transport coefficients}

The transmission matrix $T(\varepsilon)$ determines the transport coefficients of interest, like electrical conductance, heat conductance, and thermopower.
In the linear response regime these coefficients can be expressed in terms of the parameters
$L_{n}$ ($n=0,1,2$)  defined as \cite{sivan,mahan-book}
\begin{eqnarray}\label{Eq:int}
L_{n}=-\frac{1}{h}\int  d\varepsilon
(\varepsilon-\mu)^n\frac{\partial
f}{\partial\varepsilon}T(\varepsilon).
\end{eqnarray}
Thus, the charge conductance, $G$, is given by the formula
\begin{eqnarray}\label{Eq:9}
G=e^2 L_{0}.
\end{eqnarray}
Note, this formula for the conductance is equivalent to the Landauer formula, which for zero temperature reads
$G=(e^{2}/\hbar )T(\varepsilon=\varepsilon_{F})$.
In turn, the electronic contribution to the thermal conductance, $\kappa_e$, can be written as
\begin{eqnarray}\label{Eq:11}
\kappa_e=\frac{1}{T}\left(L_{2}-\frac{
L_{1}^2} {L_{0}}\right).
\end{eqnarray}
The Seebeck coefficient (or thermopower) $S$ is defined by the voltage
drop $\delta V$ generated by the temperature difference $\delta T$ as $S =
-\delta V/\delta T$, calculated for vanishing charge current, $J = 0$.
As a result the Seebeck coefficient may be written as  \cite{sivan,mahan-book}
\begin{eqnarray}\label{Eq:Seeb}
S\equiv-\left[\frac{\delta  V}{\delta
T}\right]_{J=0}=-\frac{1}{eT}\frac{
L_{1}}{L_{0}},
\end{eqnarray}
where $-e$ is the electron charge ($e>0$).
The above formulas are valid when no spin accumulation can built up in the system due to fast spin relaxation. Below they will be used to calculate
numerically the transmission coefficient and also the transport parameters.

The situations is different when the spin  relaxation in the leads is slow, so spin accumulation may occur. Writing $\delta V_\sigma=\delta V +\hat{\sigma}\delta V^s$ for the spin dependent voltage, where $\delta V$ is the electrical voltage while $\delta V^s$ is the spin voltage, one can define charge thermopower, $S =-\delta V /\delta T$, and also the
spin thermopower, $S^s =-\delta V^s /\delta T $. As the former  thermopower describes the electrical voltage $\delta V$ induced by a temperature gradient, the latter
one describes the thermally-induced spin voltage $\delta V^s$.
The spin and electrical  thermopowers can be calculated from the formulas \cite{dubiPRB09,swirkowicz}
\begin{eqnarray}\label{Eq:Ss_ac}
S^s =-\frac{1}{2eT}\left(\frac{L_{1\uparrow}}{
L_{0\uparrow}}-\frac{L_{1\downarrow}}{ L_{0\downarrow}}\right),
\end{eqnarray}
\begin{eqnarray}\label{Eq:S_ac}
S =-\frac{1}{2eT}\left(\frac{L_{1\uparrow}}{
L_{0\uparrow}}+\frac{L_{1\downarrow}}{ L_{0\downarrow}}\right).
\end{eqnarray}
In turn, the electronic contribution to the heat conductance
in this transport regime is then given by the formula \cite{swirkowicz}
\begin{eqnarray}\label{Eq:kappa_ac}
\kappa_e=
\frac{1}{T}\sum_\sigma\left(L_{2\sigma}-\frac{L_{1\sigma}^2}{L_{0\sigma}}\right).
\end{eqnarray}
The functions $L_{n\sigma}$ ($n=0,1,2$)  are defined as in Eq.~(\ref{Eq:int}) but with $T_\sigma (\varepsilon)$ instead of $T(\varepsilon)$.
Note that all the above formulas for the spin and charge thermopowers as well as the heat conductance
apply to the situation when the temperature in the leads is the same in both spin channels.
In a general case, however, the temperature of the electron reservoirs may be spin-dependent, too.

\section{Numerical results}

Now, we present numerical results for the general situation with ferromagnetic electrodes, $0<p<1$. The limiting case of half-metallic ferromagnetic leads, corresponding to $p=1$, is not interesting from the point of view of Kondo physics as the corresponding Kondo temperature goes to zero in the parallel magnetic configuration considered here. One can show that in the corresponding two-level model, one of the levels is decoupled from the leads, see Appendix. Accordingly, no Kondo screening can take place for $p=1$.

We begin with the transmission function, $T(\varepsilon )$, and the local density of states (DOS) for the quantum dot, $\rho(\varepsilon)=-(1/\pi ){\rm Im}\{G^{r}\}$, in the absence as well as in the presence of
external magnetic field. Then we analyze transport parameters including linear electrical conductance, $G$,
electronic contribution to the heat
conductance, $\kappa_{e}$, thermopower, $S$, and the resulting parameter of thermoelectric
efficiency, $ZT$. We consider separately the situations with short spin diffusion length in external leads (no spin accumulation and no spin thermoelectric effects) and
the case when spin thermoelectric effects appear.

\subsection{Transmission and spectral functions}

\begin{figure*}
\includegraphics[scale=0.6]{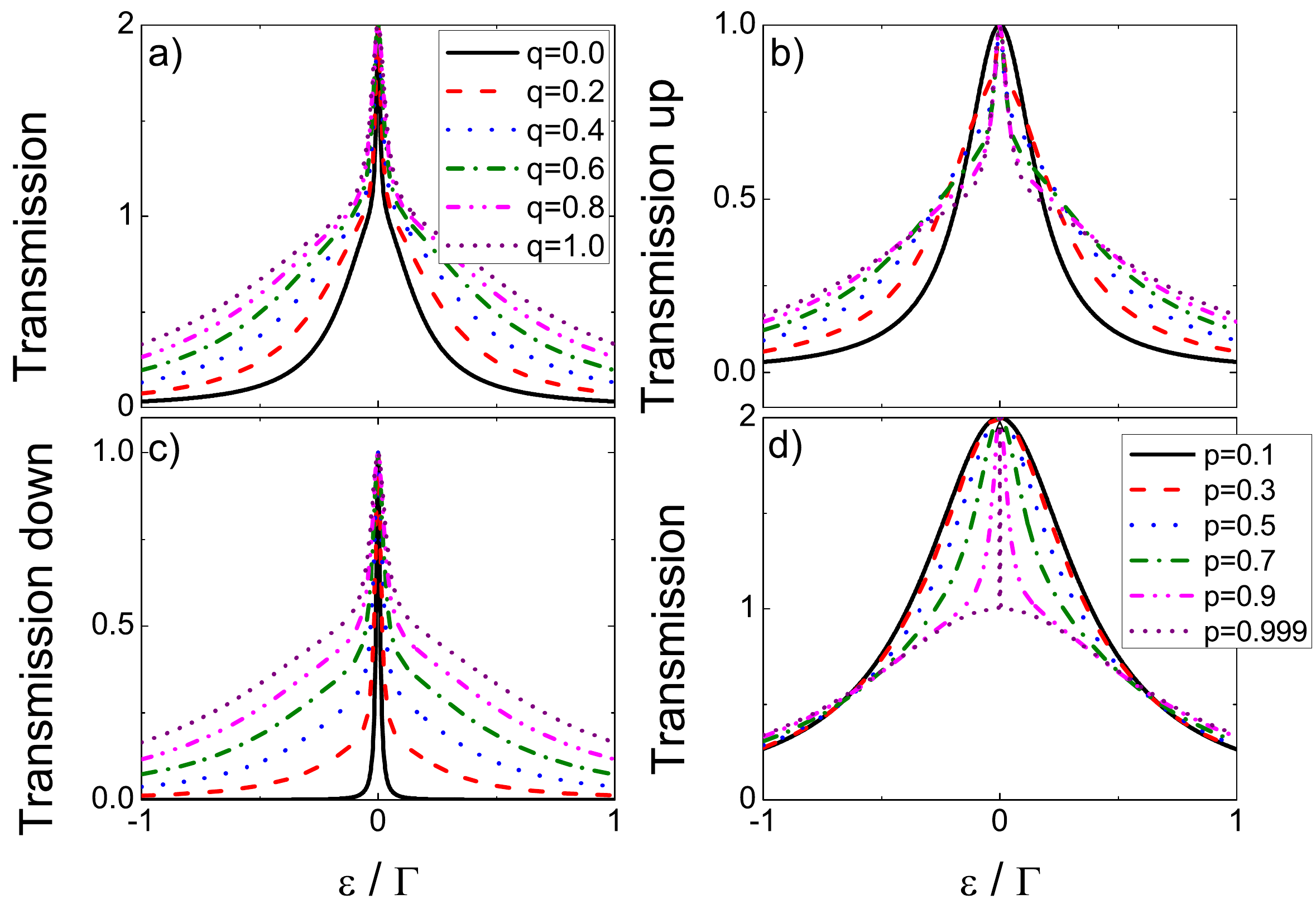}
\caption{\label{Fig:1} Total transmission (a) and its spin-up (b) and spin-down (c) components, calculated for  $p=0.9$ and indicated values of the parameter $q$. (d) Total transmission for $q=1$ and indicated values of the polarization $p$. The other parameters are: $B=0$, $\varepsilon_d=-3\Gamma$, $U=6\Gamma$, $k_BT=0$.}
\end{figure*}

Let us begin with the case of zero  magnetic field, $B=0$, when the bare dot level is spin degenerate. We consider first symmetric Anderson model, $\varepsilon_d=-U/2$. The slave boson technique takes into account spin fluctuations in the dot. Due to these spin fluctuations, the dot's spin is screened and a Kondo cloud develops in the leads. This results in the Kondo resonance  in the dot's density of states, and consequently in the Kondo effect in transport characteristics. We note that the spin fluctuations are coherent when  tunneling amplitudes for electrons of both spin orientations are sufficiently large. Tunneling with Rashba spin-orbit interaction also contributes to the spin fluctuations in the dot. The resonances in the density of states lead to a nonzero transmission through the otherwise Coulomb blocked quantum dot.
Due to the Rashba spin-orbit interaction, the two spin channels are not independent, which leads to quantum interference effects, that are absent in the limit of independent spin channels.

In figure~\ref{Fig:1}(a-c) we show the zero-temperature transmission function for different values of the  Rashba interaction strength (tuned by the parameter $q$), and for relatively large spin polarization in the ferromagnetic leads, corresponding to $p=0.9$. Figure 1a shows the total transmission function, while the corresponding spin components are presented in figure 1b and figure 1c. Since there are two spin channels in the electrodes, the maximum transmission function is equal to 2. When the Rashba interaction is absent ($q=0$), the transmission for spin-up electrons has Lorentzian shape and the peak is relatively broad. In turn, the transmission peak for spin-down electrons is very narrow. In both cases maximum transmission is equal to unity. The total transmission has then the shape which results from the  superposition of one broad and one narrow peaks. When the Rashba interaction is switched on, the total transmission function  through the quantum dot becomes broader due to the  additional tunneling channel with an effective flip of the electron spin in the dot~\cite{orellanaNano}.  Note, that the transmission  for spin-up electrons has now not a simple Lorenzian shape, as it was for $q=0$, but this shape is more complex.

In figure 1d we show the total transmission for $q=1$ and different values of the polarization factor $p$. When $p$ changes from $p=0$ to the limit of  $p=1$, but $p<1$, the shape of the total transmission changes from Lorentizan at small values of $p$ to the shape  typical of the Dicke effect for $p$ close to 1. More precisely, the peak for $p$ close to 1  consists of one broad and one narrow peaks which are superimposed at the Fermi level. The broad peak corresponds to spin-up orientation while the narrow peak is associated with spin-down orientation.
As in the original Dicke effect~\cite{dicke1}, one may associate the narrow
central peak in the transmission with a subradiant  state (weakly coupled to the electrodes), while the broad peak with a superradiant state (strongly coupled to the electrodes). The narrow peak in transmission becomes broader with decreasing spin polarization of the leads, and the Dicke line shape of the transmission transforms into a Lorenzian one. This resembles the effect of indirect hopping in a double quantum dot structure~\cite{trochaJNN,trochaJPCM}.

\begin{figure*}
\includegraphics[scale=0.85]{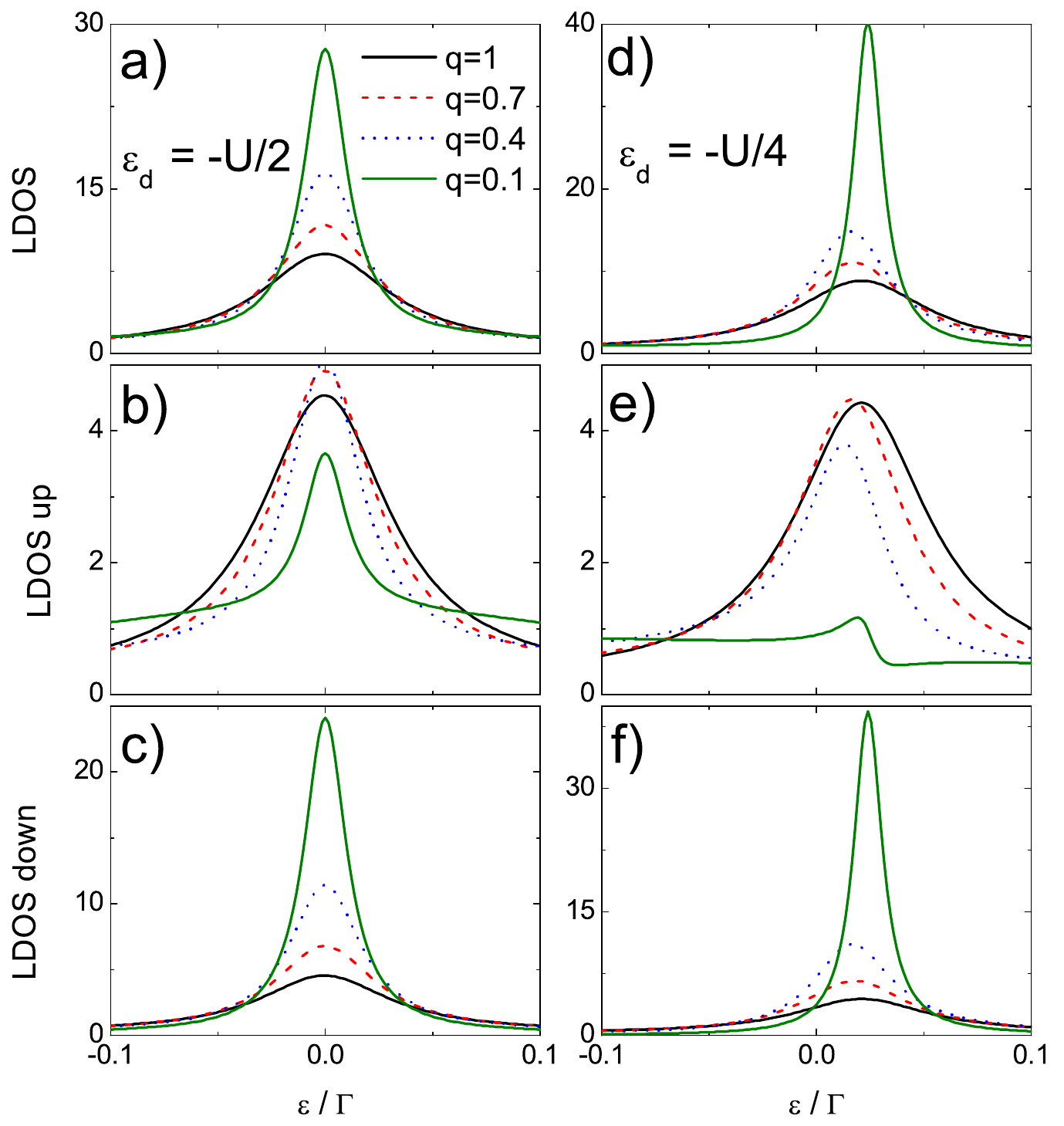}
\caption{\label{Fig:2}Local density of states (a) and its spin-up (b) and spin-down (c) components calculated for indicated values of the parameter q and for the symmetric model ($\varepsilon_d=-U/2$). Parts  d), e), and f) present the same as in the corresponding parts a), b) and c) but for the asymmetric case ($\varepsilon_d=-U/4$). The other parameters: $p=0.9$, $B=0$, $U=6\Gamma$, and $k_BT=0$.}
\end{figure*}

Coupling of the dot to both electrodes results in renormalization of the dot's level energy.
It is  already well known that spin-conserving tunneling processes between the dot and ferromagnetic electrodes generate an effective exchange field which may split the dot's level energy \cite{martinek}. In a particle-hole symmetry point, $\varepsilon_d =-U/2$, the effective exchange field, however, disappears. Spin-flip tunneling due to Rashba spin-orbit coupling contributes to the energy renormalization and also leads to mixing of the spin states in the dot. Thus, the renormalization and splitting of the dot's level is a result of the interplay of effects due to spin-flip (Rashba) and spin-conserving tunneling processes.
The energy renormalization and spin mixing of the dot's states has a significant influence on the Kondo resonance, which is an interference effect of various cotunneling processes with effective flip of the dot's spin.
Local density of states (LDOS) corresponding to the Kondo resonance depends then on the renormalized dot's states.
In figure 2(a,b,c) we show the density of dot's states associated with the Kondo peak for a symmetric model, $\varepsilon_d =-U/2$. One finds then a single Kondo peak at the Fermi level, which depends on the Rashba coupling parameter $q$.
Strictly speaking, Rashba interaction diminishes hight of the total Kondo peak, see figure 2(a), where the peak hight decreases with increasing $q$. In turn, spin components of the LDOS depend on the Rashba coupling in a more complex way, i.e. the Rashba coupling magnifies hight of the spin-up Kondo peak (figure 2(b)) and diminishes hight of the spin-down peak (figure 2(c)).

The situation is different in asymmetric situation, in which the dot is described by an asymmetric Anderson model, $\varepsilon_d\ne -U/2$. The corresponding LDOS for $\varepsilon_d\ne -U/4$ is shown in figure \ref{Fig:2}(d,e,f). When $q=1$, the LDOS has the shape of a Lorentzian peak for both spin orientations, and the peaks are shifted above the Fermi level and have comparable hight. When $q$ decreases, the spin-up LDOS is reduced (see figure 2(e)) while the spin-down LDOS is increased (see figure 2(f)).

\begin{figure*}[t]
\begin{centering}
\includegraphics[scale=0.8]{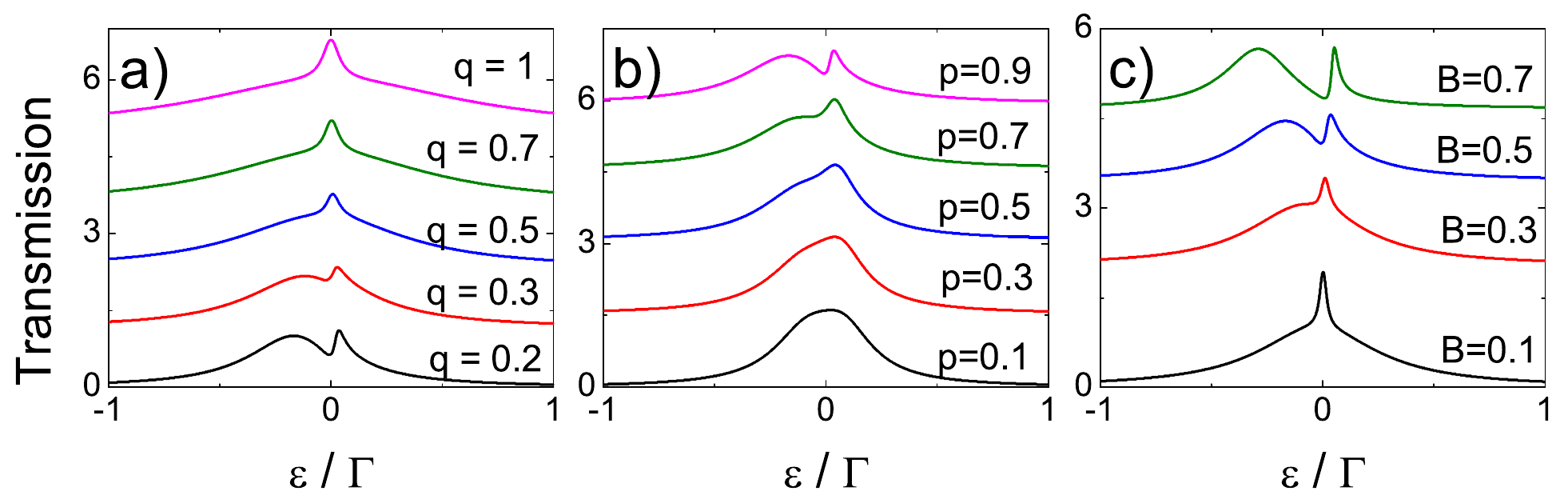}
\par\end{centering}
\caption{\label{Fig:3} Evolution of the Fano antiresonance in the transmission shown as a  function of energy measured from the Fermi level. (a) Transmission for indicated parameter $q$ and $p=0.9$, $B=0.5\Gamma$; (b) Transmission for indicated polarization factor $p$, and  $q=0.2$, $B=0.5\Gamma$; (c) Transmission for indicated values of the magnetic field $B$, and $p=0.9$, $q=0.2$. The other parameters: $U=6\Gamma$, $\varepsilon_d=-3\Gamma$, $k_BT=0$.  Each consecutive curve is shifted up to increase visibility. }
\end{figure*}

Now, we include magnetic field, which lifts  spin degeneracy of the bare dot's level,  and investigate its influence on the picture described above.
In figure~\ref{Fig:3}(a) the total transmission function $T(\varepsilon)$ for a symmetric Anderson model is displayed for different values of the parameter $q$, while $p$ and $B$ are kept constant, $p=0.9$ and $B=0.5\Gamma$. For $q=1$ there is only one peak which is located at the Fermi level. This indicates that the Rashba interaction suppresses the effects due to magnetic field. As the parameter $q$ decreases, the role of Rashba coupling becomes diminished, so the influence  of magnetic field effectively increases. The peak becomes then split into broad and narrow peaks and a dip develops at the Fermi level. The total transmission reveals then a typical antiresonance behavior at the Fermi level (see the curve for $q=0.2$ in figure 3(a).
This Fano-like antiresonance originates from destructive quantum interference of electron waves transmitted through the broad and narrow states.
When now the polarization factor $p$ decreases, this antiresonance behavior disappears, as shown in figure 3(b). Note, the curve for $q=0.2$ in figure 3(a) coincides with the curve for $p=0.9$ in figure 3(b). For $p=0.1$ only a broad peak remains in the transmission.  Similarly, when the external field  decreases, the antiresonance
disappears as well, as shown in figure 3(c). Note, the curve for $p=0.9$ in figure 3(b) coincides with the curve for $B=0.5$ in figure 3(c). For small values of  magnetic field, the transmission function resembles the Dicke line shape, see the curve for $B=0.1$ in figure 3(c)

\subsection{Thermoelectric effects: no spin accumulation}

Now, we analyze the thermoelectric phenomena in the system under consideration. The basic thermoelectric characteristics, like the Seebeck coefficient, electron and heat conductance, and figure of merit have been calculated in the linear response regime and in the limit of zero spin accumulation in the leads. The role of spin accumulation will be analyzed in the following subsection. When calculating the thermoelectric effects we assume temperature below the corresponding Kondo
temperature, $T<T_K$ . The latter was estimated numerically by calculating the electrical conductance $G$ as a function of temperature, and the Kondo temperature was then taken as the temperature at which the conductance falls down to half of its maximum value, as shown in figure 4. From this estimate follows that $k_BT_K/\Gamma \approx 0.1$ for $q=0$, and increases with increasing  $q$.

\begin{figure}
\includegraphics[scale=0.3]{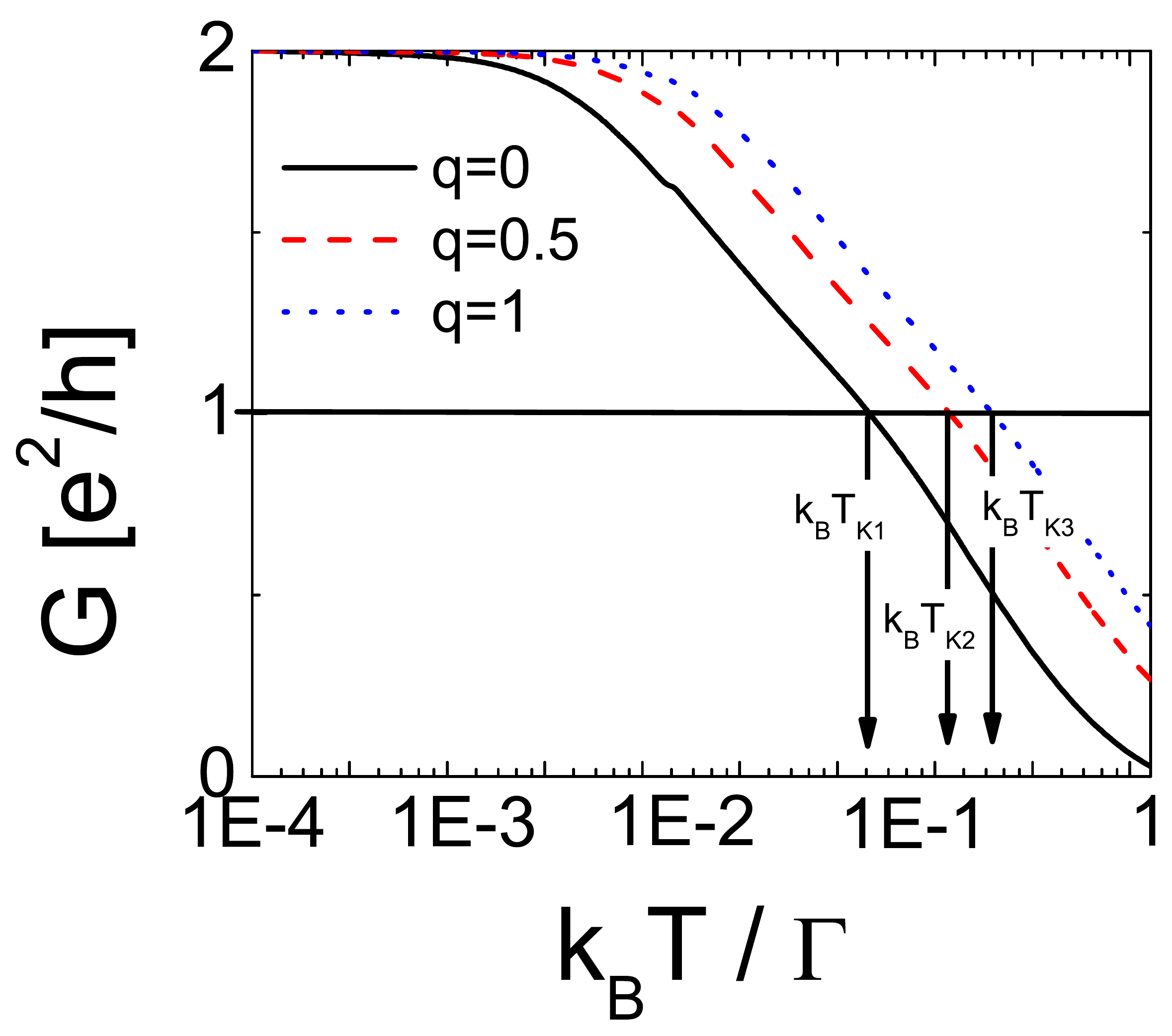}
\caption{\label{Fig:4} Electrical conductance $G$ as a function of temperature for indicated values of the parameter $q$.
The other parameters: $B=0$ and $U = 6\Gamma$. The arrows indicate the Kondo temperatures $T_K$ corresponding to the three values of $q$.}
\end{figure}

\begin{figure*}
\includegraphics[scale=0.46]{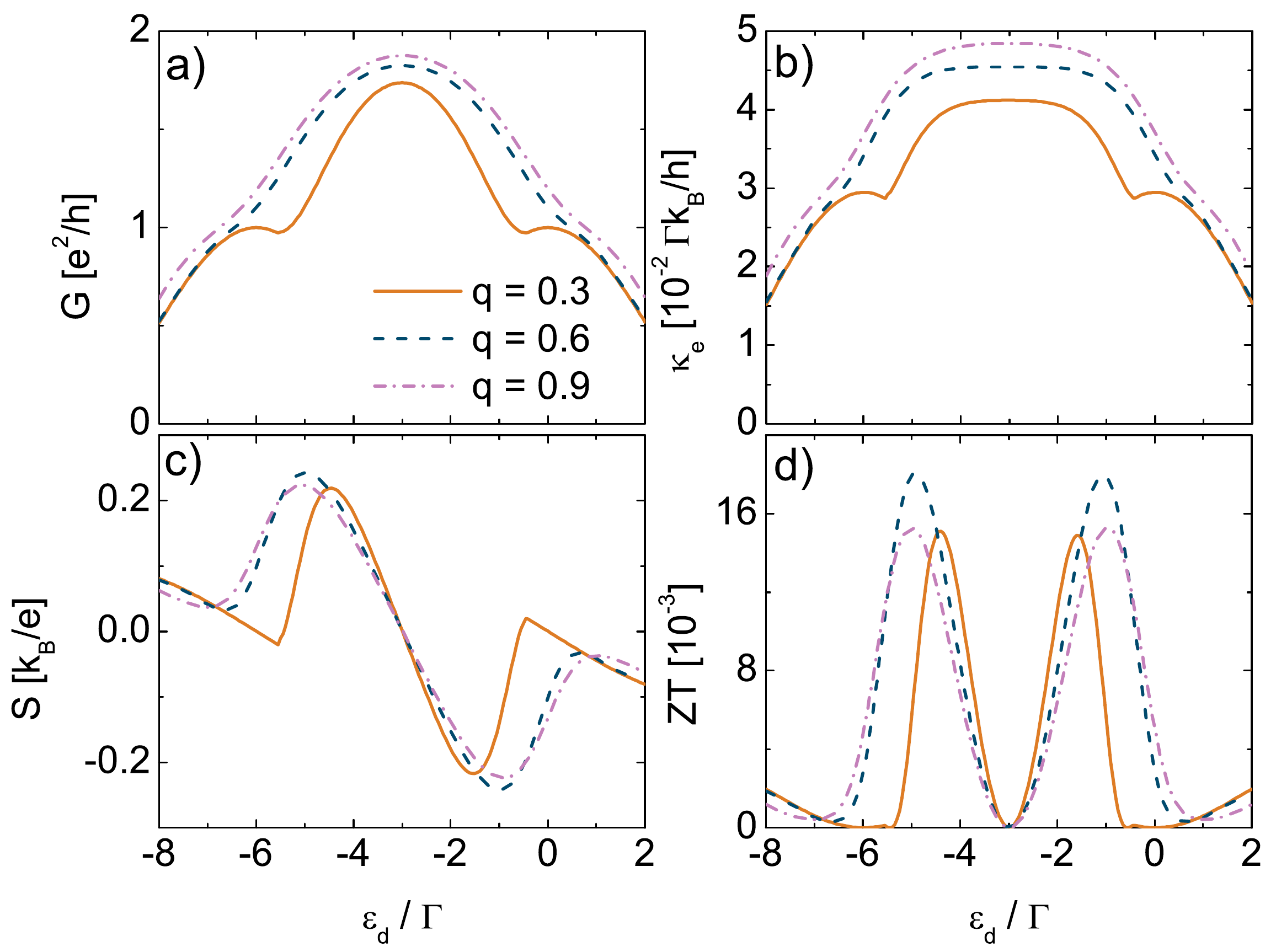}
\caption{\label{Fig:5} Electrical conductance $G$ (a),
heat conductance $\kappa_e$ (b), thermopower $S$ (c), and  figure of merit $ZT$ (d), calculated as a function of the dot's level energy
for indicated values of the parameter $q$.
The other parameters: $B=0$, $U = 6\Gamma$, $k_BT=0.009\Gamma$.}
\end{figure*}

The transport and thermoelectric coefficients of the system in the absence of magnetic field are shown in
figure~\ref{Fig:5}. These coefficients are plotted there as a function of the dot's energy level $\varepsilon_d$
for different values of the parameter $q$. Consider first the electrical conductance, figure~\ref{Fig:5}(a). According to figure~\ref{Fig:1}, the transmission function $T(\varepsilon)$ in the symmetrical Anderson model reveals two peaks; one narrow and one broad, with the corresponding maxima located at the Fermi level. For  asymmetrical position of the dot level, $\varepsilon_d\ne -U/2$, the maxima in spin dependent LDOS and consequently also in transmission functions   are shifted away from the Fermi level, as follows from figure~\ref{Fig:2}(d-f). Thus, the conductance reaches maximum in the particle-hole symmetry point, $\varepsilon_d=-U/2$, as expected, and is  reduced in asymmetric situations. Due to a nonzero temperature assumed in figure~\ref{Fig:5} (but below the Kondo temperature), the conductance maximum is below the quantum limit $2e^2/h$.
When increasing the magnitude of the parameter $q$, which enhances the Kondo correlations, one observes an increase in the conductance towards the quantum limit. In turn, for small values of $q$, the role of Kondo correlations is reduced, so the central Kondo peak in the conductance is also reduced and the resonances corresponding to $\varepsilon_d$ and $\varepsilon_d+U$ become well resolved.

Electronic contribution to the thermal conductance is shown in figure~\ref{Fig:5}(b). Since the temperature assumed in figure~\ref{Fig:5} is low,
this contribution, as a function of the dot level position, reveals similar behavior as the corresponding electrical conductance. The only qualitative difference is that the maxima in thermal conductance are flat, contrary to those in the electrical conductance.

The Seebeck coefficient, shown in  figure~\ref{Fig:5}(c), vanishes in the particle-hole symmetry point, $\varepsilon_d=-U/2$, for any value of the parameter $q$. This is a consequence of the symmetrical density of states around the Fermi level ($\varepsilon=0$) at this position of the dot level. The thermally-induced charge current carried by electrons is then fully compensated by the charge current due to holes, which results in zero net charge current, and therefore zero voltage drop. When the dot's level is slightly above the symmetric point, $\varepsilon_d>-U/2$, the thermopower becomes negative as the dominant contribution to the current comes from electrons due to a shift of the resonances in LDOS above the Fermi level, see figure~\ref{Fig:2}(d-f).
In turn, when the dot's level is slightly below the particle-hole symmetry point,
the main contribution to current comes from holes and the thermopower is positive.
Since the thermopower is an antisymmetric function of the dot's level position measured with respect to the symmetric point, we will restrict further discussion  to the case of $\varepsilon_d < -U/2$. Thus, when $\varepsilon_d$ decreases starting from $\varepsilon_d = -U/2$, the thermopower $S$ increases (starting from $S=0$) until some maximum value of $S$ is reached. Then, the thermopower  decreases with a further decrease in $\varepsilon_d$ and reaches a minimum at a certain value of $\varepsilon_d$. When $\varepsilon_d$  decreases further, $S$ starts to grow again. For small values of the spin-orbit parameter $q$, the Seebeck coefficient changes sign twice close to the above mentioned point where $S$ has a local minimum.

\begin{figure*}
\includegraphics[scale=0.46]{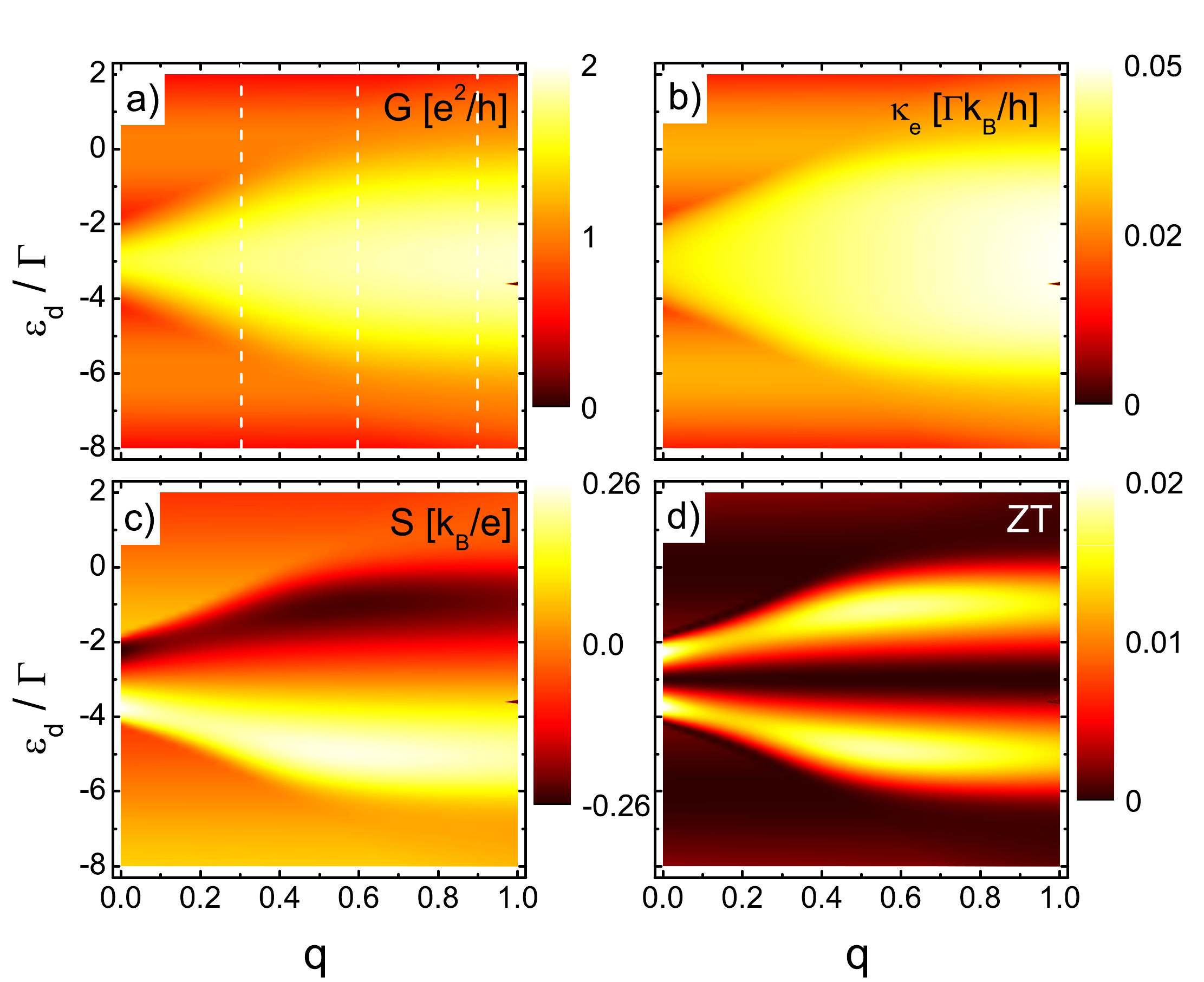}
\caption{\label{Fig:6} Electrical conductance $G$ (a),
heat conductance $\kappa_e$ (b), thermopower $S$ (c), and  figure of merit $ZT$ (d), calculated as a function of the dot's level energy
and the parameter $q$. The dashed lines represent cross-sections plotted in
figure 4.
The other parameters: $p=0.9$, $B=0$, $U = 6\Gamma$, and $k_BT=0.009\Gamma$.}
\end{figure*}

The thermoelectric efficiency measured by the dimensionless figure of merit $ZT$ is displayed in figure~\ref{Fig:5}(d) as a function of the dot's level energy. For all values of $q$, the figure of merit $ZT$ vanishes in the particle-hole symmetry point, where also the thermopower vanishes. Magnitude of $ZT$ grows rather fast when the dot level moves away from the symmetric point, and $ZT$ reaches maxima at the points where contributions from the Kondo peaks become suppressed. For small values of $q$, $ZT$ also vanishes at the points where the corresponding thermopower changes  sign (compare figures~\ref{Fig:5}(c) and \ref{Fig:5}(d)). The magnitude of $ZT$, however, is rather small, indicating that thermoelectric efficiency in the Kondo regime, though it can be controlled and modified by the Rashba spin-orbit interaction, is rather negligible from the practical point of view.

The density plots presenting the electrical conductance and thermoelectric coefficient as a function of dot's level energy and the parameter $q$ are shown in figure 6. The explicit dependence on $q$ is especially interesting as it shows directly the influence of the Rashba interaction. Note, figure 5 presents cross-sections of figure 6 for specific values of the parameter $q$, as indicated by the dashed lines in figure 6(a).

\begin{figure*}
\includegraphics[scale=0.46]{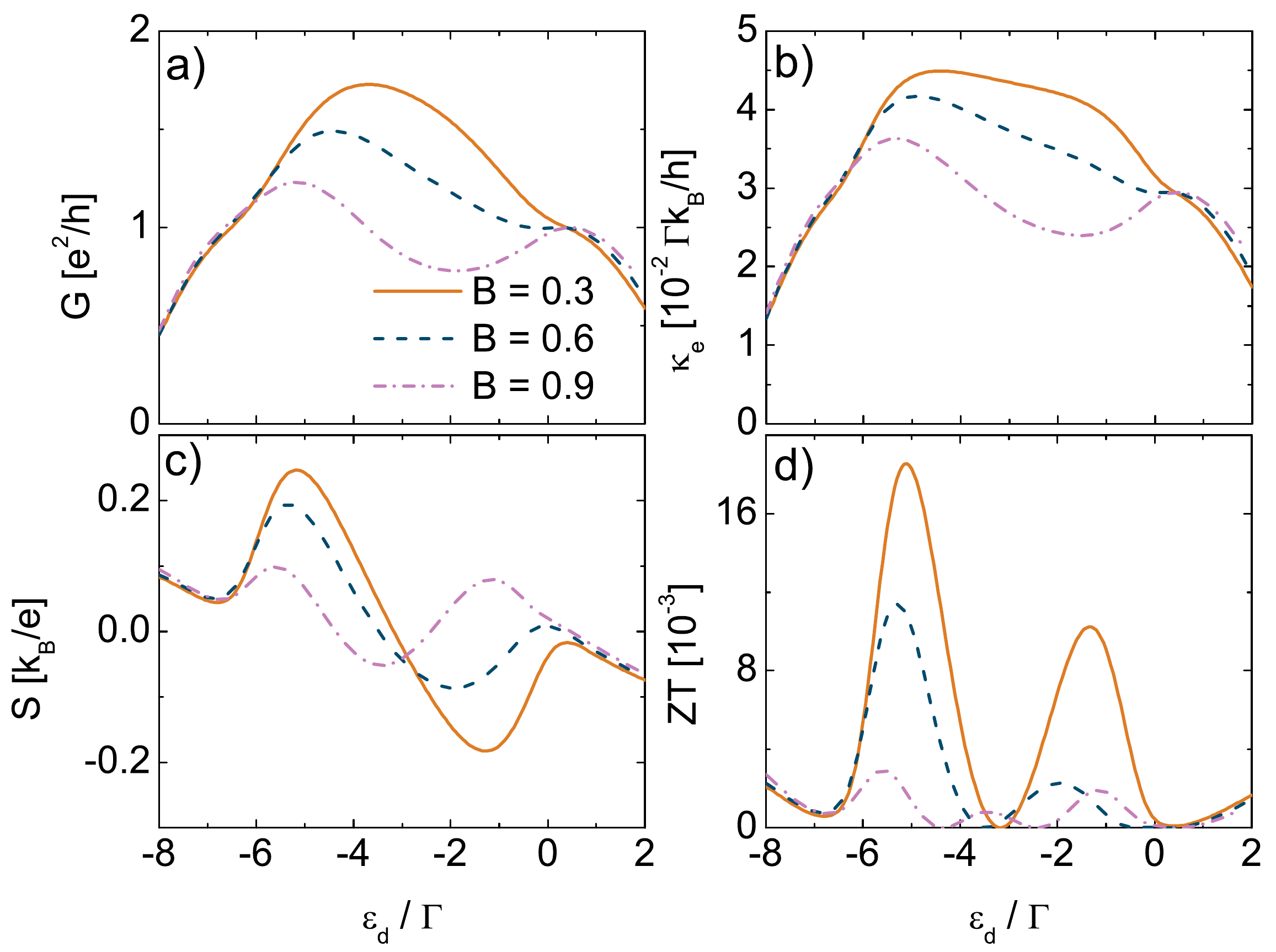}
\caption{\label{Fig:7} Electrical conductance $G$ (a),
heat conductance $\kappa_e$ (b), thermopower $S$ (c), and figure of merit $ZT$ (d), calculated as a function of the dot's level energy
for indicated values of the magnetic field $B$. The other parameters: $p=0.9$, $U = 6\Gamma$, $k_BT=0.009\Gamma$, and $q=0.6$.}
\end{figure*}

The thermoelectric properties of the system can be also controlled by an external  magnetic field.  Figure~\ref{Fig:7} displays the transport and thermoelectric characteristics calculated for $p=0.9$,  $q=0.6$, and indicated values of $B$. Note, magnetic field breaks the particle-whole symmetry which is present in the case of $B=0$. For  small magnetic fields, the electrical conductance, figure~\ref{Fig:7}(a), reveals a broad peak and achieves maximum close to the particle-hole symmetry point. A sufficiently large magnetic field leads to splitting of the Kondo peak in LDOS, which results in a decrease of the zero bias Kondo anomaly. The electrical conductance in the Kondo regime becomes then suppressed, and for a sufficiently large magnetic field only the side-bands corresponding to the resonances at $\varepsilon_d=0$ and $\varepsilon_d=-U$ survive. This suppression is clearly visible in figure~\ref{Fig:7}(a) for $B/\Gamma =0.9$.

Thermal conductance, shown in figure~\ref{Fig:7}(b) as a function of the dot's level energy and indicated values of magnetic field,  behaves in a similar way as the electric conductance, though the maximum for small magnetic fields is broader than in the electrical conductance, as already mentioned before. This similarity in qualitative behavior usually appears at low temperatures.

The thermopower is shown in figure~\ref{Fig:7}(c). It is interesting to note, that thermopower in the Kondo regime changes sign with increasing magnetic field.  Accordingly, the figure of merit $ZT$ is also reduced with increasing magnetic field, as shown in figure~\ref{Fig:7}(d). Interestingly,  $ZT$ is then significantly larger than in the case shown in figure~\ref{Fig:5}(b).

\begin{figure*}
\includegraphics[scale=0.46]{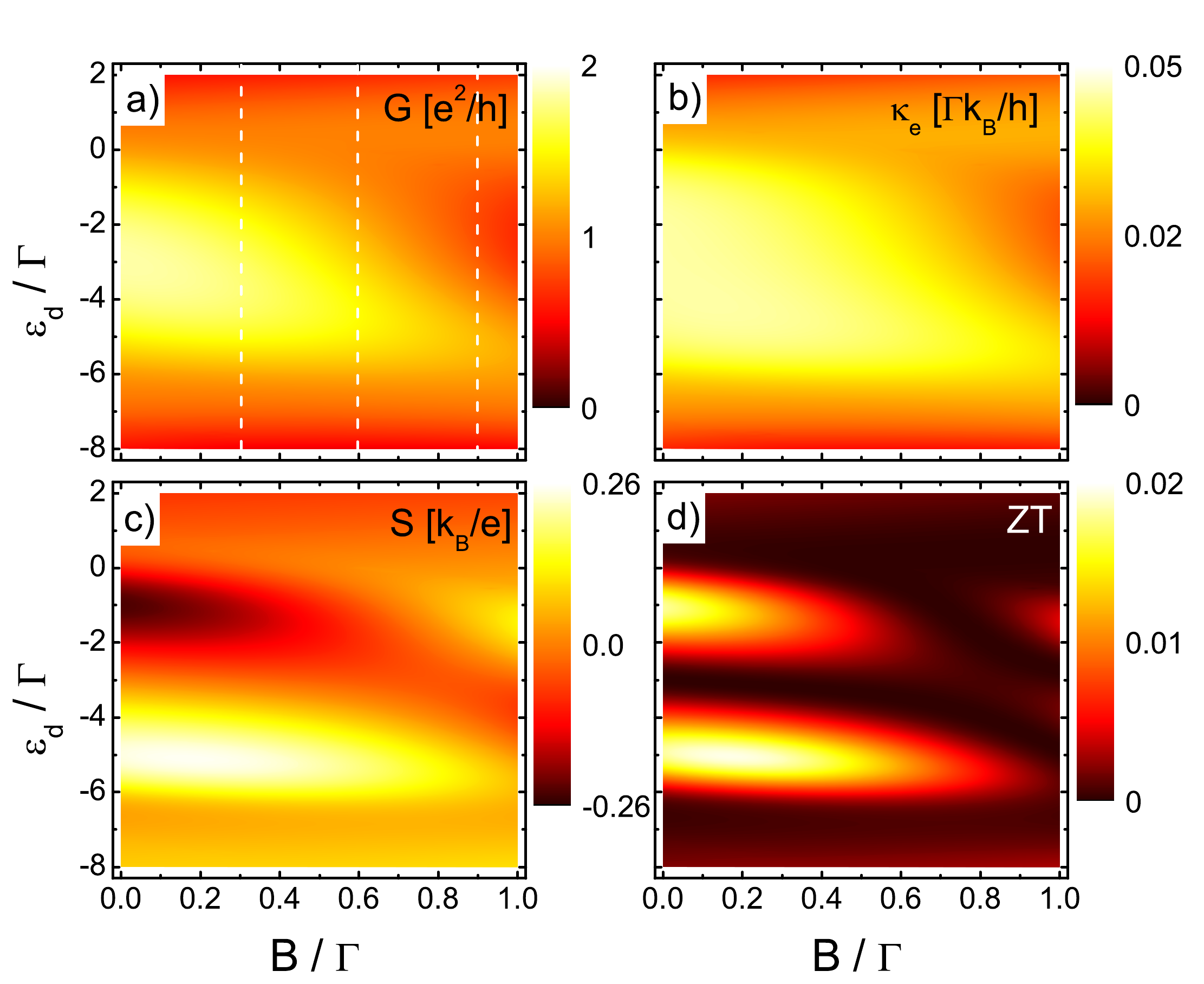}
\caption{\label{Fig:8} Electrical onductance $G$ (a),
heat conductance $\kappa_e$ (b), thermopower $S$ (c),  and figure of merit $ZT$ (d), calculated as a function of the dot's level energy and magnetic field $B$.
Dashed lines in (a) represent cross-sections plotted
in figure 7. The other parameters: $U = 6\Gamma$, $k_BT=0.009\Gamma$, $p=0.9$, and $q=0.6$.}
\end{figure*}

The influence of magnetic field on the results presented above is shown explicitly in figure~\ref{Fig:8}, where the dependence of the transport and thermoelectric coefficients on the dot's level position and magnetic field is presented for $p=0.9$ and  $q=0.6$. Note, figure 6 presents cross sections of figure 8 for magnetic fields indicated by dashed lines in figure 8(a).

\subsection{Spin thermoelectric effects}

In this subsection we present some numerical results in the situation when spin relaxation processes in electrodes are very slow so their role can be neglected. In such a case, a spin accumulation builds in the leads. The following discussion will be restricted to the thermopower. First, one should note that electric thermopower studied in the preceding sections is modified in the limit of slow spin relaxation (presence of spin accumulation) in comparison to what was found in the limit of strong spin relaxation (no spin accumulation). Second, an additional thermopower can be now defined, so-called spin thermopower (spin Seebeck effect), which is one of the most pronounced spin related thermoelectric phenomena.

\begin{figure*}
\includegraphics[scale=0.6]{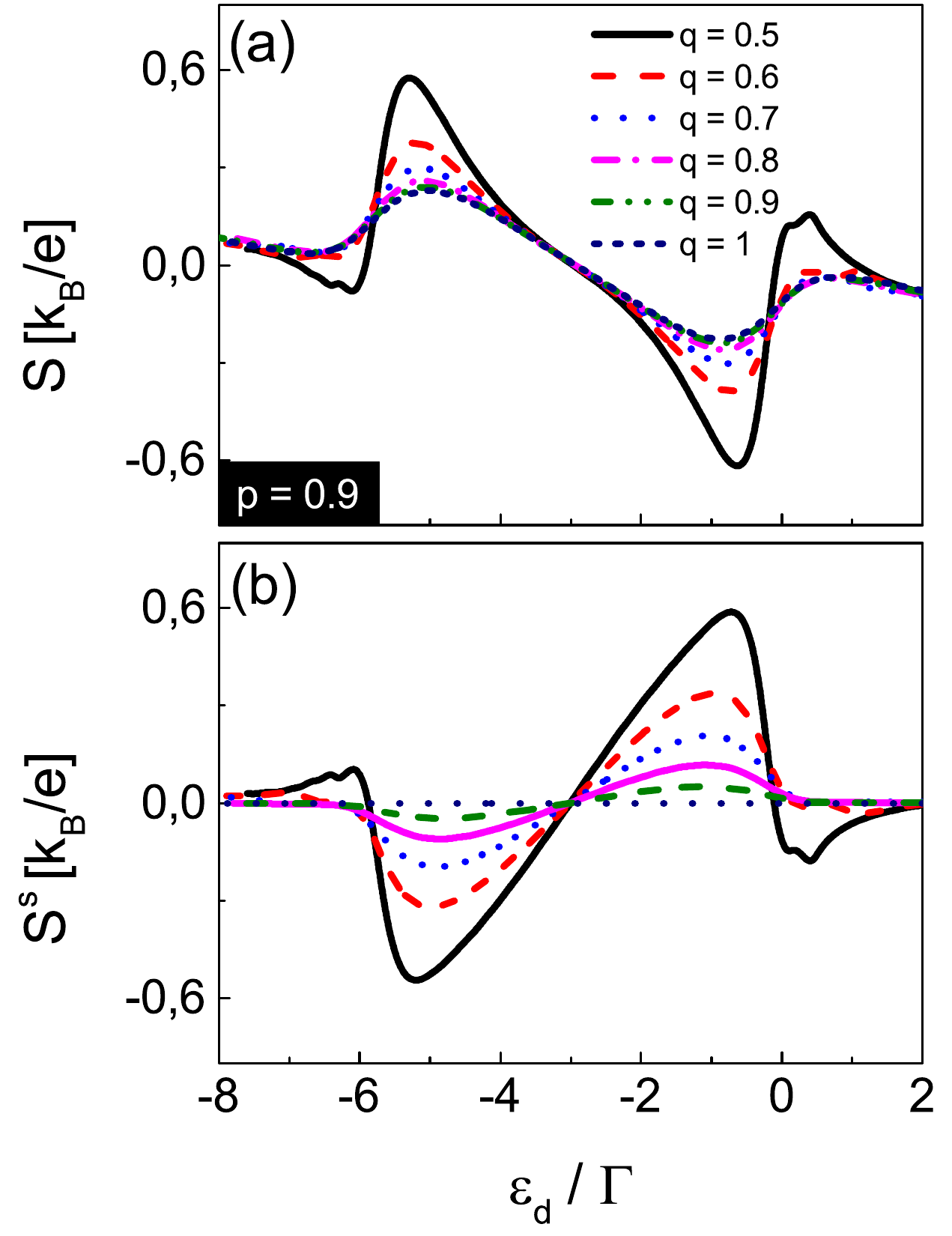}
\caption{\label{Fig:9} Thermopower (a) and spin thermopower (b) as function of the dot's level energy for indicated values of the parameter $q$. The other parameters: $U = 6\Gamma$, $T=0.009\Gamma$, $B=0$.}
\end{figure*}

In figure~\ref{Fig:9} we show both electric thermopower (see figure~\ref{Fig:9}(a)) as well the spin thermopower (see figure~\ref{Fig:9}(b)) for several values of the parameter $q$. Both thermopowers are shown there as a function of the dot's level energy. The spin thermopower has opposite sign when compared to the electric thermopower. Otherwise, it depends on the level position in a similar way as the electric thermopower. It vanishes in the particle-hole symmetric point as well as at the resonances.  The absolute magnitudes of both thermopowers are also comparable, although there are some quantitative differences. Both $S$ and $S^s$ are reduced with increasing $q$. However, as the electrical thermopower remains nonzero at $q=1$, the corresponding spin thermopower vanishes due to mixing of both spin channels caused by the spin-orbit interaction (see the curve for $q=1$ in figure~\ref{Fig:9}(b). Thus, from the point of view of spin thermopower, the spin-orbit interaction is rather undesired.

\section{Summary and conclusions}

We have analyzed thermoelectric effects in a quantum dot connected to two ferromagnetic leads. Apart from direct spin-conserving tunneling between the dot and leads, we have also included tunneling with spin-orbit coupling of Rashba type. The considerations have been restricted to the low temperature regime, where the Kondo correlations play a crucial role in electronic transport. To calculate transport properties we used the slave boson technique in the mean field approximation.

Owing to the interplay of spin conserving and Rashba type tunneling processes, we observed both Dicke and Fano like interference effects in the transmission function. These effects modify electrical conductance as well as the thermoelectric parameters. Numerical results have been presented for two limiting situations. The first situation concerns the case with when spin relaxation processes  exclude spin accumulation in the leads and therefore exclude spin thermopower. The second one, in turn, refers to the opposite situation, when the spin relaxation in the leads is slow. One can observe then a nonzero spin thermopower. The latter describes spin voltage generated by a temperature gradient. The  spin thermopower, however, becomes reduced by the spin-orbit Rashba interaction.

\begin{acknowledgments}
This work was supported by the National Science Center in Poland as the Project No. DEC-2012/04/A/ST3/00372.
\end{acknowledgments}

\appendix
\setcounter{section}{1}
\section*{Appendix: Unitary transformation}

The  initial model, see hamiltonian~\ref{hamiltonian}, is equivalent to  a spinless two-level quantum dot in an effective magnetic field~\cite{kashcheyevs, silvestrov}. Assuming for simplicity that both the spin-conserving and spin-nonconserving tunneling amplitudes are independent of {\bf k}  and electrode index, $V_{\mathbf{k}\beta\sigma}=V_{\sigma}$, $V_{\mathbf{k}\beta\sigma}^{so}=V_{\sigma}^{so}=\sqrt{q}V_{\sigma}$), we write the corresponding transformation in the form
\begin{eqnarray}
\begin{pmatrix}d_{1}^{(\dagger)}\\
d_{2}^{(\dagger)}
\end{pmatrix}=\frac{1}{\sqrt{1+q}}\begin{pmatrix}1 & -i\sqrt{q}\\
-i\sqrt{q} & 1
\end{pmatrix}\begin{pmatrix}d_{\uparrow}^{(\dagger)}\\
d_{\downarrow}^{(\dagger)}
\end{pmatrix},
\end{eqnarray}
where the new operators obey the fermionic anticommutation relations,
$\left\lbrace d_{i},d_{j}^{\dagger}\right\rbrace = \delta_{ij}$   and $\left\lbrace d_{i}^{(\dagger)},d_{j}^{(\dagger)}\right\rbrace = 0$ for $i,j=1,2$.

The transformed hamiltonian of the system can be written as
\begin{eqnarray}
H'=H_{e}+H_{d}'+H_{t}',
\end{eqnarray}
where $H_{e}$ is left unchanged, while the dot hamiltonian $H_{d}'$ and tunneling hamiltonian $H_{t}'$ assume the following forms:
\begin{eqnarray}
H_{d}' & = & \sum_{i=1,2}\varepsilon_{i}d_{i}^{\dagger}d_{i}-B_{y}\hat{S}_{y} + Ud_{1}^{\dagger}d_{1}d_{2}^{\dagger}d_{2},\\
H_{t}' & = & \sum_{\mathbf{k}\beta}\left(V_{\uparrow}'c_{\mathbf{k}\beta\uparrow}^{\dagger}d_{1}+V_{\downarrow}'c_{\mathbf{k}\beta\downarrow}^{\dagger}d_{2}+h.c.\right),
\end{eqnarray}
where
\begin{eqnarray}
\varepsilon_{i}& = &\varepsilon_{d}\mp\frac{B(1-q)}{2(1+q)},\\
B_{y} & = & \frac{2B\sqrt{q}}{1+q},\\
V_{\sigma}' & = & \sqrt{1+q}V_{\sigma}.
\end{eqnarray}
In Eq.(A3), $\hat{S}_{y}=(1/2)\sum_{ij}(\sigma_{y})_{ij}d_{i}^{\dagger}d_{j}$ is the $y$ component of a pseudospin operator. The renormalization of the spinless energy levels $\varepsilon_{i}$ depends on the effective field resulting from the interplay of the Rashba term and external magnetic field. Specifically, the field dependence is completely suppressed for $q=1$, so the states related to energy levels $\varepsilon_{i}$ are degenerate. Similar situation appears in the case of $B=0$.

The transformed hamiltonian describing tunneling between the electrodes and the dot is diagonal in the spin space, 
\begin{eqnarray}
\mathbf{\Gamma}^\prime =\begin{bmatrix}\Gamma_{\uparrow}^\prime  & 0\\
0 & \Gamma_{\downarrow}^\prime 
\end{bmatrix}=\begin{bmatrix}(1+p)(1+q)\Gamma & 0\\
0 & (1-p)(1+q)\Gamma
\end{bmatrix}.
\end{eqnarray}
The above formula clearly shows that one of the states becomes decoupled from the leads for half-metallic electrodes, i.e. for $p=1$. Accordingly, no Kondo state  can be then formed.

\end{document}